\documentclass[preprint,11pt,numbers,times]{elsarticle}
\usepackage{titlesec}
\titleformat*{\subsection}{\bfseries}
\titleformat*{\section}{\large\bfseries}
\titleformat{\paragraph}
    {\normalfont\itshape}
    {\theparagraph}{}{}
    
\usepackage{amssymb}
\usepackage{lineno}

\usepackage{url}
\usepackage{todonotes}
\usepackage{natbib}
\setcitestyle{square,comma,sort&compress}
\usepackage{xcolor}
\usepackage{booktabs}
\usepackage{changepage}

\journal{journal}

\begin{document}

\begin{frontmatter}

\title{Open government geospatial data on buildings for planning sustainable and resilient cities}

\author[doa,dre]{Filip Biljecki\corref{cor1}}

\affiliation[doa]{organization={Department of Architecture, National University of Singapore},%
            country={Singapore}}
\cortext[cor1]{Corresponding author at: Department of Architecture, School of Design and Environment, National University of Singapore, 4 Architecture Dr, 117566 Singapore}

\author[geog]{Lawrence Zheng Xiong Chew}

\affiliation[dre]{organization={Department of Real Estate, National University of Singapore},%
            country={Singapore}}

\affiliation[geog]{organization={Department of Geography, National University of Singapore},%
            country={Singapore}}

\author[mcc,tub]{Nikola Milojevic-Dupont}

\author[mcc,tub]{Felix Creutzig}

\affiliation[mcc]{organization={Mercator Research Institute on Global Commons and Climate Change},%
            country={Germany}}

\affiliation[tub]{organization={Technical University Berlin},%
            country={Germany}}

\begin{abstract}
As buildings are central to the social and environmental sustainability of human settlements, high-quality geospatial data are necessary to support their management and planning. 
Authorities around the world are increasingly collecting and releasing such data openly, but these are mostly disconnected initiatives, making it challenging for users to fully leverage their potential for urban sustainability.
We conduct a global study of 2D geospatial data on buildings that are released by governments for free access, ranging from individual cities to whole countries.
We identify and benchmark more than 140 releases from 28 countries containing above 100 million buildings, based on five dimensions: accessibility, richness, data quality, harmonisation, and relationships with other actors.
We find that much building data released by governments is valuable for spatial analyses, but there are large disparities among them and not all instances are of high quality, harmonised, and rich in descriptive information.
Our study also compares authoritative data to OpenStreetMap, a crowdsourced counterpart, suggesting a mutually beneficial and complementary relationship.
\end{abstract}

\end{frontmatter}

\section{Introduction}\label{sc:introduction}

\begin{figure}[h]
    \centering
    \includegraphics[width=0.9\textwidth]{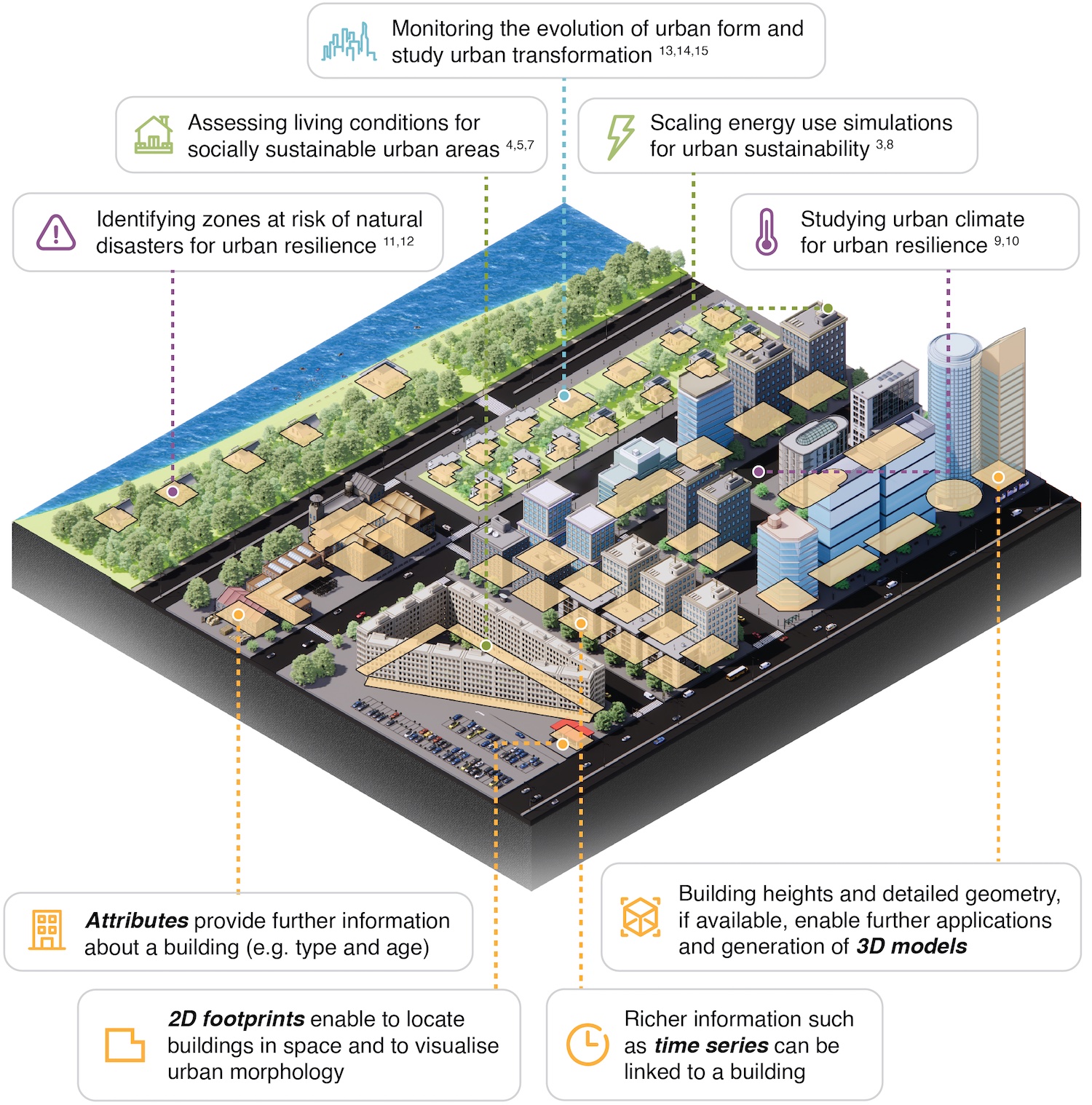}
    \caption{\textbf{Geospatial data on buildings provides key information for urban sustainability, resilience, and transformations.} 2D buildings footprint (in light orange) can provide important localised information about the building stock and spatial structure of cities, relevant to answer an array of research questions of relevance to answer pressing questions about cities in twenty-first century. Further semantic information and 3D information offer opportunities for richer analyses by enabling to differentiate the building stock at a finer level and to study the vertical dimension of cities.}
    \label{fig:illustration}
\end{figure}

\noindent Buildings are a central infrastructure in urban environments, providing social services such as shelter that are key for human societies.
Thus, establishing and maintaining maps of buildings has been an important activity for city and national governments for decades to support various governance processes (e.g.\ urban planning and cadaster).
Openly accessible geospatial data on buildings underpin knowledge creation for urban sustainability, resilience and transformations~\citep{elmqvist2019sustainability}, as illustrated in  Figure~\ref{fig:illustration}. Such big data on cities establish a quality transition in research and potentially in governance on urban sustainability because street- and building specific opportunities emerge that previously were subject to high transaction costs; with the use of big and hybrid data context-specific solutions become accessible city-wide \citep{creutzig2019upscaling}. 

The large portfolio of use cases includes, for urban sustainability, energy~\citep{Roth.2020}, urban farming~\citep{Palliwal.2021}, socio-economic studies~\citep{Feldmeyer.2020}, routing~\citep{Wang.20181ia}, noise~\citep{Stoter:2020db} or solar simulations~\citep{Zhang.2019}; for urban resilience microclimate~\citep{Yuan.2020}, thermal comfort~\citep{Gamero-Salinas.2020}, flood exposure~\citep{Huang.20201dd} or disaster management~\citep{Westrope.2014}, and for urban transformations, urban morphology~\citep{Hecht:2019jd}, population~\citep{Schug.2021} or greenery~\citep{Wu.2020}. 
To recast the boundaries of research of the role of buildings for planetary sustainability, it is essential that fine-grained, high-quality, and comparable data are available across the world. Achieving such an integration in a timely manner requires cooperation between the different actors and a convergence towards best practices and standards.

Digital data on buildings come in different flavours, from aggregated statistics on a national level to geospatial datasets characterising the building form to architecturally detailed datasets~\citep{Milojevic-Dupont.2020b,Chong.2021,Biljecki.2021}. The two main types of structured data on individual buildings at large scale are geospatial vector data and tabular data. Both can be combined, but that is often not necessarily the case. Geospatial vector data can include the footprint of the ground surface of buildings as 2D maps or 3D geometries with different levels of detail, possibly representing architectural details from the envelope and indoor components. Other geospatial data include street view imagery, satellite imagery, and point clouds, but these usually require to be processed and matched to other existing datasets. Tabular data may include attributes such as the address, type, age or construction material of the building, but also time series, e.g.\ energy use. While such data is essential for a portfolio of use cases, aggregated statistics at coarser levels (e.g.\ district, city, region or country) are often the only available snapshot on building stock characteristics.

Governments from the city to the state level (e.g.\ through their national mapping agencies) have been the traditional actor developing and maintaining authoritative spatial data on buildings, primarily for cadastral and topographic purposes~\citep{Dorn.2015,Du.2017}, and now they are increasingly open to the public. 
In the last decades, other majors actors have started complementing authorities, namely volunteered geoinformation (VGI) initiatives, companies and research institutions~\citep{Miller.2020}.
In particular, OpenStreetMap (OSM) --- as the key instance of VGI --- contains more than half billion buildings around the world, including low-income regions~\citep{Biljecki.2020,Yeboah.2021}, and in some cases with high completeness and quality~\citep{Hecht:2013jl,Brovelli:2018eh}.
Thanks to its crowdsourcing and collaborative nature, OSM is inherently open and represents an alternative when governments do not release building data~\citep{Li2020:os}. %
Some commercial companies are producing comprehensive maps of building stocks, but mainly for online map services or purchase, and with only some areas offered as open data, e.g.\ Microsoft's dataset of building footprints in the United States (\url{https://www.microsoft.com/en-us/maps/building-footprints}). %
Research institutions, universities, and international organisations~\citep{Miller.2020} have also generated highly valuable datasets~\citep{Wu.2020,Dukai.2020}, for example, through large-scale remote sensing analysis~\citep{Esch_2017}.

Notwithstanding the emergence of other prominent actors, governments remain a key producer of spatial data on buildings, but there is no large-scale comparative study on open government data on buildings. Little is known about data availability across regions, as well as trends and challenges in term of data quality, completeness, and relationships with users and other producers of building data. Such knowledge could both help governments identify best practices and facilitate the usage by a broader range of stakeholders. The scarce existing literature indicates that even in the same country, building information is often not harmonised (e.g.\ list of data attributes to be collected), inhibiting comparative analyses, data exchange, and development of tools~\citep{Malhotra.2020}. More literature exists on open geospatial data in general~\citep{Quarati.2021,Donker.2016,Johnson.2017,Vancauwenberghe.2018,Mulder.2020}, and open government data in cities~\citep{Sayogo.2014,Dawes.2016}. Studies~\citep{Zuiderwijk.2014} and initiatives such as the Global Open Data Index (\url{https://index.okfn.org/}) and the Open Data Barometer (\url{https://opendatabarometer.org/}) also developed instruments to benchmark data quality.

In this paper, we conduct a global comparative study on open geospatial datasets on buildings released by local, regional, and national governments on all continents. We focus on the most common spatial data on buildings: their footprints represented as 2D geometry~\citep{McGlinn.2021}, optionally accompanied by information such as age, type or height, represented as attributes. 

Our key contribution is gathering and analysing the largest inventory of open datasets worldwide to date. We identified more than 140 jurisdictions in 28 countries around the world releasing open data on more than 100 million buildings. We then analyse the datasets cutting across five main dimensions: accessibility, richness, data quality, harmonisation, and relationship with other actors, then assess them through a set of thirteen quantitative and qualitative features, e.g.\ frequency of updates, usage of standards or condition of access. Finally, we discuss the implications of these results for governments, other producers of building data and users, focusing on understanding the role of government data amid the increasing landscape of high-quality building data, especially OpenStreetMap. %
This study is part of a continuous aggregation effort available as a website (\url{https://ual.sg/project/ogbd/}).

\section{Results}\label{sc:results}

\subsection{Dimension 1: Accessibility}

    \medskip
    \paragraph{Existence}
    
\begin{figure}[h!]
    \centering
    \includegraphics[width=\textwidth]{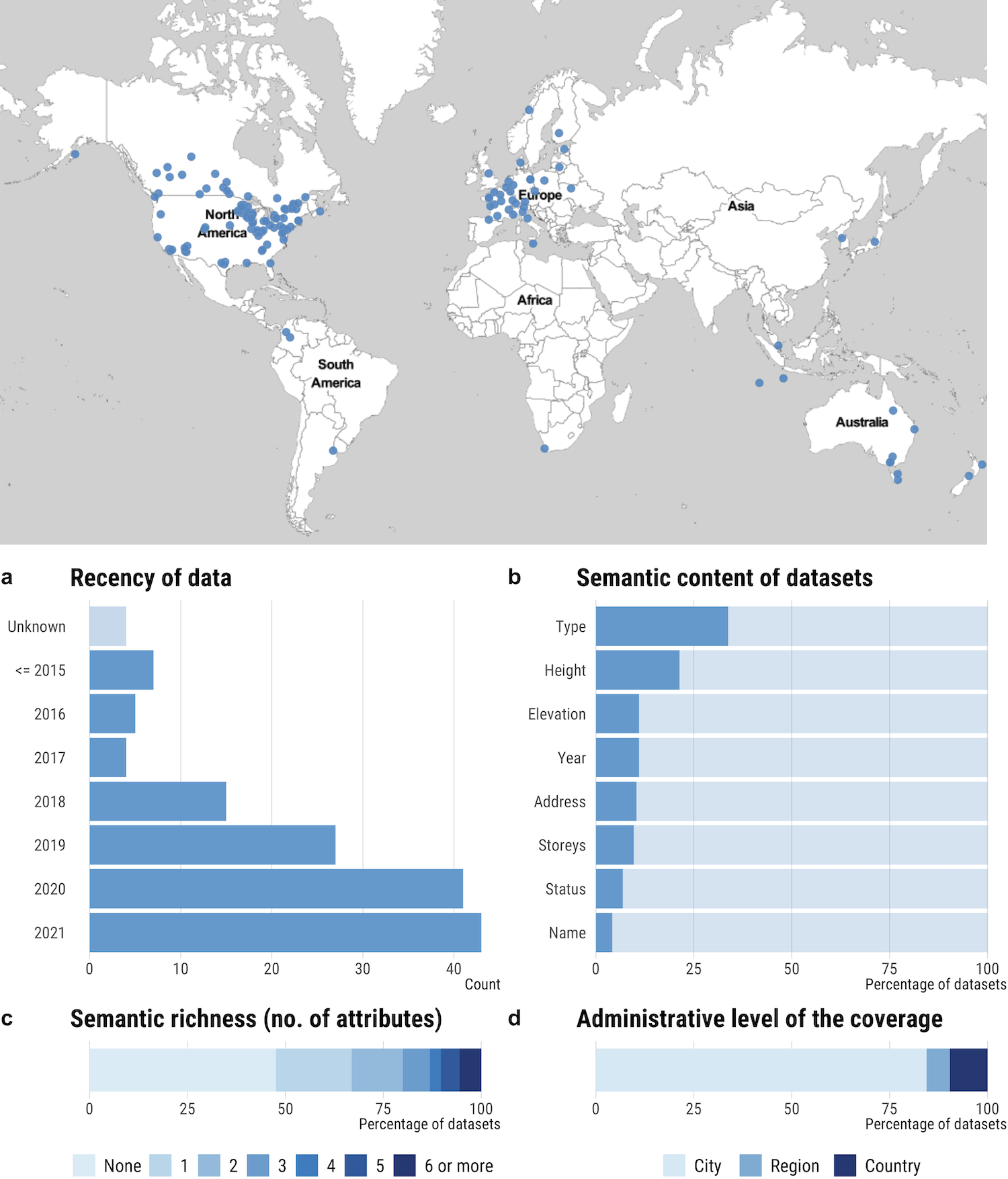}
    \caption{\textbf{Main spatial and quantitative patterns of open government data on buildings.} Map: locations where open government data was found; (a) year of release or last update of the datasets; (b) the most frequently available semantic attributes in the datasets and their prevalence; (c) administrative scope of the datasets; and (d) number of semantic attributes available in a dataset.}
    \label{fig:summary_metrics}
\end{figure}

    We identified datasets on all inhabited continents. The majority of the datasets is found in high-income countries, predominantly from North America and Europe (Figure~\ref{fig:summary_metrics}).
    Very few datasets are from low- and middle-income economies in Africa, Asia or South America, e.g.\ from Columbia or South Africa. Several high-income countries fail to release data, indicating cultural or administrative barriers. In many countries, only a few, usually large, cities are covered, indicating independent initiatives by resource-rich local governments.

    \paragraph{Openness}
    
    Most datasets are fully accessible without cost, registration, or enquiry.
    However, access to data is not always without issues, and various practices in term of openness and access to the data make the collection of the data challenging and time consuming.
    Barriers to access include broken links, invalid files, and complex APIs --- diminishing accessibility to stakeholders without advanced technical skills. 
    For example, certain download services e.g. Web Feature Service (WFS) are primarily meant to download small local samples and may require convoluted workflows to download large areas. 
    Further, some datasets advertised on open data portals turn out not to be open data; a common practice is to display data in a web map without download options.

    \paragraph{International visibility}
    
    International visibility is important for global or comparative studies including multiple geographies.
    In non-English speaking countries, open data portals are rarely available in English, inhibiting the identification of data for comparative studies.
    Another impediment are attributes, metadata and documentation available only in local language(s).
    While it is understandable that governments publish the data according to the official language(s) of the country, these datasets may be left out of international comparative studies due to unintelligibility.

\subsection{Dimension 2: Richness}

    \medskip
    \paragraph{Spatial extent and completeness}
    
    There tends to be an order of magnitude more datasets at the city level than on regional or national levels (Figure~\ref{fig:summary_metrics}). Datasets at the national levels seem to be currently only existing in Europe, where there has been several important releases in the last year, with the exception of Japan and New Zealand.
    In several instances, datasets do not cover all buildings in the jurisdiction, which may not be indicated in the metadata.
    For example, some datasets have large swaths of land entirely omitted, while some include only buildings of a particular class or with certain criteria (e.g.\ only commercial buildings or those with the largest footprint areas).

    \paragraph{Presence of attributes}

    About half of the datasets (53\%) only inform about the 2D geometry of buildings, whereas
    the other 47\% contains at least one attribute about buildings %
    (Figure~\ref{fig:summary_metrics}). The most common attribute (about one third of datasets) informs on the type of the building. Building height, elevation, address, year of construction, and number of storeys follow as additional less frequent attributes. The relatively large share (28\%) of datasets containing either height or storey information indicates a potential for generating 3D city models using extrusion~\citep{Biljecki.2020}.
    A few datasets, e.g. in France and Finland, go beyond the most common attributes to include information on building characteristics such as roof type and amenities including electricity, water supply, heating system, elevator, etc.

    \paragraph{Level of detail}

    Most datasets contain polygons representing the building footprint, but in a few cases, especially in low-income countries, only coordinates or addresses were provided, which are of limited use in spatial analyses. In some cases, different levels of detail are available (e.g.\ points and polygons) with the lower level of detail available for free, while the higher is available at a cost. The footprints may contain more or less architectural details (see Figure~\ref{fig:osm}). Numerical attributes can also have different level of precision, due to aggregation of different datasets with different estimation methods. This may be indicated in the data, e.g.\ for building heights in France, where the altimetric precision is given. Categorical attributes are also far from being harmonised and may have quite different granularity. For example, building type may contains only few classes (e.g.\ public/commercial/residential), but also 10+ classes (e.g.\ school, museum, hospital, etc.).

\subsection{Dimension 3: Quality}

    \medskip
    \paragraph{Availability of essential metadata}
    
    Nearly all (more than 90\%) datasets provide information on the year of last update.
    However, due to the difference in website configurations, as well as user experience/interface design (UI/UX) formats, it often takes time to locate where updates and acquisition methods are described.
    License status is also often not straightforward. Sometimes, there are links provided to government-specific data sharing rules due to the sensitivity of the data. A number of the datasets (28\%) is openly available under the Creative Commons License for sharing. The metadata may not be conclusive in describing what certain attributes mean.

    \paragraph{Frequency of updates}
    
    Half of the datasets has been either created or last updated within the last one year (Figure~\ref{fig:summary_metrics}). Practices regarding updates vary dramatically. On the one hand, there are instances released a decade ago with no update since then. Further, some metadata claim yearly updates, while we were not able to find evidence of this. On the other hand, some datasets are updated on a weekly basis, e.g.\ Estonia and Netherlands. However, we have found very few instances of versioning, e.g.\ in New York City, enabling to study urban development.
    
    \paragraph{Validity of geometries}
    
    Validating the geometric content of the datasets reveals that they are virtually free of geometric errors. Less than 0.1\% of buildings have invalid geometry and more than half of the datasets do not have a single error.
    Such quality is in stark difference from 3D datasets on buildings, in which the share of errors is often more than an order of magnitude higher~\citep{Biljecki:2019wh}.
    Nevertheless, geometric consistency does not warrant the positional accuracy of the data.

\subsection{Dimension 4: Harmonisation}

    \medskip
    \paragraph{Standard identifiable}
    
    Although there are efforts to develop building information standards, using international standards is not a common practice for 2D building datasets, limited to a small share of the datasets encountered. The main example is the INSPIRE standard, which led to harmonisation of practices in the EU on dimensions such as metadata, licences, download methods, or data encoding. 
    While some data are modelled according to national standards, the lack of international standards impacts interoperability, as it forces users aiming to perform comparative studies to verify the above-mentioned dimensions in each dataset.

    \paragraph{Formats}
    
    About 86\% of the datasets are available in ESRI shapefile, and the remaining datasets are stored in more than five different formats. 
    While widely supported, shapefile is a proprietary format. 
    Some datasets are available in custom formats, greatly reducing accessibility. 
    40\% of the datasets are available in two or more formats, out of which an open and non-proprietary formats are common. 

    \paragraph{Definition of a building footprint}
    Building footprints can be represented either as whole blocks, as whole buildings, as buildings parts, and in some cases several layers are provided (e.g.\ in Spain). In most cases, it appears that whole buildings are represented. %
    What defines an individual building is often unclear and may differ across regions, an issue that may cause interoperability issues and noisy results in large-scale models.

\subsection{Dimension 5: Relationship with other actors}

    \medskip
    \paragraph{Integration and relationship with OpenStreetMap}
    
\begin{figure}[h!]
    \centering
    \includegraphics[width=\textwidth]{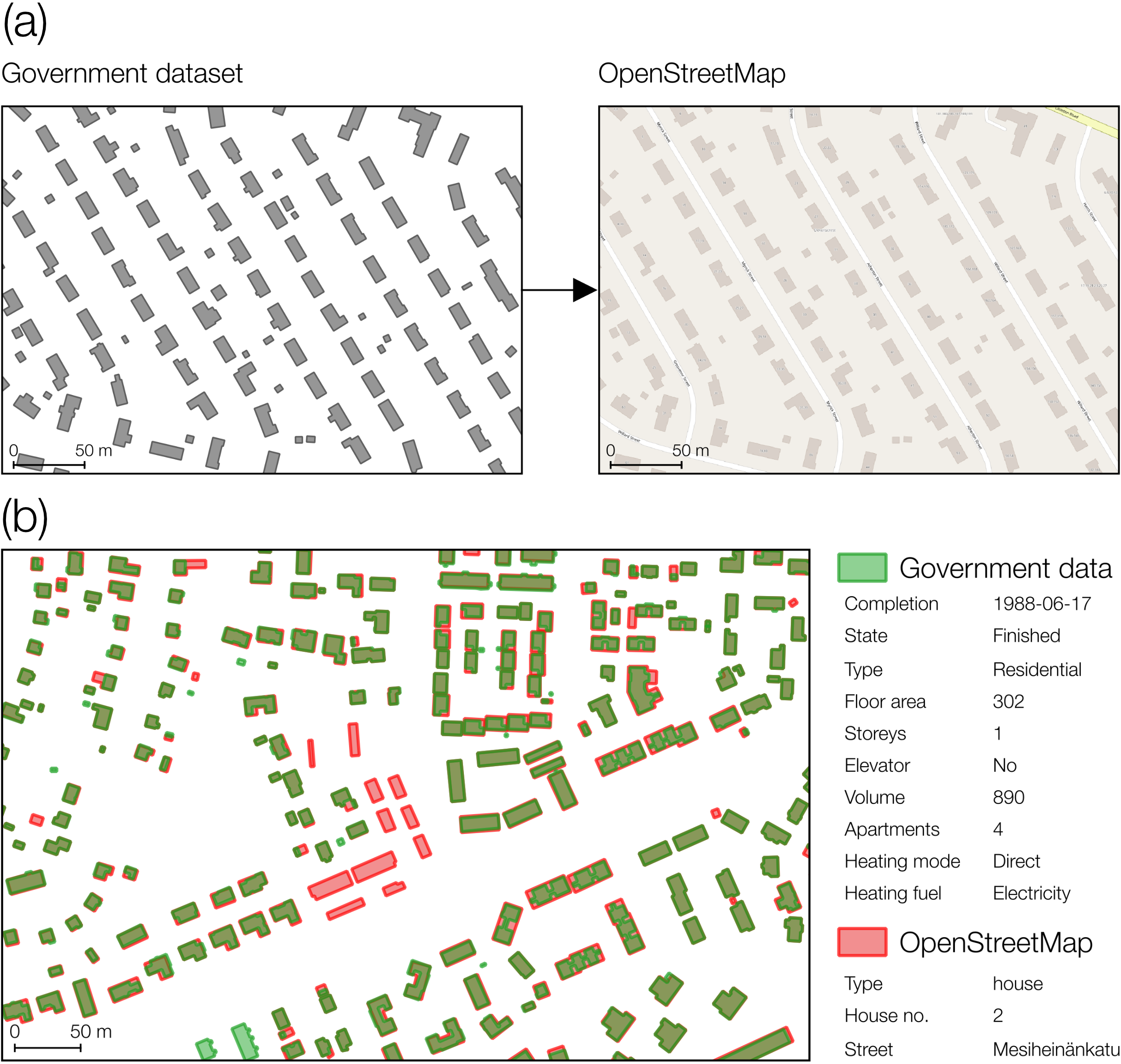}
    \caption{\textbf{Mutually beneficial relationship between open government data and OpenStreetMap.} (a) Government data can be integrated in OSM. Millions of buildings in OSM have been sourced from government data. This example shows an import of government data in OSM for an area in Middlesex County, MA, USA. 
    (b) Open government and OSM data in the same area can be complementary. This example shows both for an area in Tampere, Finland. Certain buildings are mapped in OSM but not in the government dataset (e.g.\ new buildings), and vice-versa (e.g.\ small structures such as sheds not captured by participatory mapping). Similarly, different attributes are available in either. Attribute names and values have been translated from Finnish and are exemplified for one building.
    Credits: \textcopyright\ OpenStreetMap contributors; Bureau of Geographic Information (MassGIS), Commonwealth of Massachusetts (USA), Executive Office of Technology and Security Services; Tampere region (Finland).
    }
    \label{fig:osm}
\end{figure}

    The OSM community has leveraged the open nature of many government datasets and integrated them into OSM via bulk imports.
    We find that only a quarter of government datasets we have identified has been integrated into OSM.
    Data imports usually serve well rural areas, when a dataset is released at the state or regional level, as these tend to be undermapped in OSM in comparison to urban areas~\citep{Camboim:2015gg}.
    Figure~\ref{fig:osm} shows an example where buildings in OSM has been sourced from the government data.
    Imports in OSM entail a stringent process that involves investigating potential impact of the mass import on the existing data and suitability of the licence~\citep{Zielstra.2013,Juhasz.2018}.
    It is unclear why more government datasets have not been taken advantage of. In some areas, a possible explanation is that government data, while available, is outdated, and OSM already provides data of higher quality.  
    
    \paragraph{Government data curated by a private company}
    
    In many instances, government data is distributed by a private company, e.g.\ hosted on their servers or integrated in their products.
    While it appears that such data are certainly originating from a government body, we could not find any official repository hosting the same instance, possibly hinting at cases of outsourcing, or cases that the government dataset first released by a government, with the company mirroring it on their services.
    The dataset may have been removed from their repository due to being out of date, without a newer counterpart released, and with the commercial host remaining available; or the data may not open, but has been purchased by a company.
    
    \paragraph{Involvement of universities}
    
    We find compelling cases where universities are complementing government activities with multiple purposes: standardisation, validation, and enhancement of data.
    First, they are regularly involved in developing standards and harmonisation approaches~\citep{Stoter:2020db,Biljecki.2021}.
    Second, universities provide methods and tools for validation, quality control, and benchmarking spatial data~\citep{prepair}.
    Our paper is also an example how academia can contribute to gauging the data.
    Finally, universities leverage their expertise to integrate data, multiplying their instances.
    For example, in the Netherlands, a research group has conflated multiple government datasets, and released the processed and enriched dataset as open data~\citep{Dukai.2020}.
    Further, being research-driven users of data, they are spearheading new use cases, possibly leading to influencing future data releases and enhancing data requirements.

\section{Discussion}\label{sc:discussion}

\noindent High-coverage, highly detailed, and high-quality open geospatial data on buildings are needed globally to support urban sustainability research and open governance processes, at different scales and across different application domains. An important question is how such status can be achieved most rapidly. Several actors --- governments, VGI, companies, and academic organisations --- can take part in this development, and complement each other. Our analysis demonstrates that governments are and are likely to remain a key provider of open data on buildings, but that currently the data they release are in many instances insufficient to tackle important and timely questions. See Table~\ref{tab:summary} for a summary of the results with implications and recommendations of best practices.

\subsection{Priorities from an urban sustainability perspective}
    
    \medskip
    \paragraph{Data for transforming the building stock}
    
    Buildings represent key challenges for urban sustainability, resilience, and transformations~\citep{elmqvist2019sustainability}, and new data are required to study relevant determinants and design solutions involving buildings around the world~\citep{bai2018six,zhu2019understanding,shi2016roadmap,silva2017urban,seto2017sustainability,acuto2018building,thacker2019infrastructure}. Key gaps in knowledge include the sustainability outcomes from different urban forms, the impact built-up land use on other systems~\citep{zhu2019understanding}, and how to translate this knowledge into actionable policies~\citep{zhu2019understanding,shi2016roadmap,acuto2018building}. 
    
    In this context, attaining a complete coverage of building footprints globally is a key priority, but our study indicates that open government data remain currently far from such objective. The lack of building attributes limits the range of questions that can be addressed with the data, and the number of dimensions that can be examined simultaneously~\cite{silva2017urban,seto2017sustainability}. Attributes may not be necessary for applications such as urban morphology based solely of spatial arrangements, but can substantially enrich them for social, environmental, and economic modelling. Thus, it is necessary to achieve a high coverage of basic attributes, e.g.\ building heights~\cite{zhu2019understanding} as well as increasing the availability of important by rarely available attributes such as building envelope materials. 
    
    \paragraph{Data for learning from the Global South}
    
    Well-designed buildings and neighbourhoods, as part of the critical infrastructure for basic well-being, are key to achieving multiple Sustainable Development Goals in the Global South~\citep{thacker2019infrastructure,creutzig2016urban,nagendra2018urban}. In particular, the construction and operation of new buildings in fast urbanising areas in Asia, the Middle East, and Africa are projected to consume most of remaining carbon budget in business-as-usual scenarios, thus early strategies for low-carbon developments are pivotal~\citep{creutzig2016urban,bai2018six}. Although cities in the Global South face important constraints, they also offer an overlooked capacity to innovate and experiment for sustainability~\citep{nagendra2018urban}.

    Our study reveals that there are barely any government sources from the Global South and the fastest urbanising regions, which is an important concern. Building data from the Global South is necessary, given that they exhibit distinct and statistically different issues from the Global North~\citep{nagendra2018urban}, and to enable the assimilation of narratives and local knowledge with technical data~\citep{bai2018six}. Up-to-date information and available archives are of particular importance in these regions to study urbanisation trends, see for example the NYU Atlas of Urban Expansion (\url{http://www.atlasofurbanexpansion.org/}). Finally, data on informal settlements is difficult to generate but critical~\citep{bai2018six}.

    \paragraph{Data for facilitating analyses across scales}
    
    There is a broad agreement in the urban sustainability literature on the need for more studies across scales and geographies to tackle regional and global sustainability challenges~\citep{elmqvist2021urbanization,elmqvist2019sustainability,seto2017sustainability,thacker2019infrastructure,creutzig2016urban,Milojevic-Dupont.2020}. The lack of such studies leads to various knowledge gaps. Much of current research is at the individual city level or aggregate level~\citep{elmqvist2021urbanization}, while there is a need for studies spanning multiple geographic and administrative scales. These enable to generate and act on clear targets of desired diversity at different scales~\citep{elmqvist2021urbanization}; they address the spatial scale between human and natural systems, as well as the trade-offs between social and environmental outcomes~\citep{seto2017sustainability}. 
    
    Enabling such research requires building data that is easily accessible both for the local and global user, with minimal barriers regarding data access, processing or licensing. 
    Our analysis indicates that the multiplicity of small-scale datasets with diverging practices generates substantial barriers even for experienced users. Further standardisation and consolidation of datasets is needed~\citep{thacker2019infrastructure}. The lack of data for in many world regions is also a key bottleneck for enabling comparative work~\citep{seto2017sustainability} and city typologies~\citep{baur2014urban, creutzig2015towards} that help understand common patterns and diversity, and find solutions that may transfer across contexts~\citep{thacker2019infrastructure,Milojevic-Dupont.2020}.  

\subsection{Strengthening open data strategies from governments}

    \medskip
    \paragraph{Understanding barriers}
    
    The very small number of datasets found in low- and middle-income countries seems to indicate that creating and releasing data often remains a luxury. Cultural and administrative barriers include technical, organisational or budget-related issues. These may include the cost of training staff internally or need to hire sub-contractors, difficulty to gather information that is fragmented and not necessarily digitalised across services, or concern of divulging sensitive information. In certain regions, there may not be clear perspectives of rapid progress and therefore a need for other actors to step up. But rapid progress can be observed, especially if data products are already existing, for example for internal use or commercial purposes, as this has been the case recently in the European Union with the EU directive 2019/1024, which increased pressure on governments to release data.    

    \paragraph{Implementing simple best practices}
    
    In Table~1, we provide recommendations that can substantially increase the value of the data for users. 
    Many of them are easy to implement, e.g.\ providing basic information through metadata and naming, or publicly archiving previous versions of a dataset instead of keeping only the latest version online. Some other best practices can represent a larger expenditure, e.g.\ achieving a complete coverage of key attributes, but such investments offer large societal returns~\citep{Janssen.2012,Ruijer.2020}. When providing complete datasets for attributes may be too big of an investment, recent data generation literature, e.g.\ using machine learning, also shows that partially complete datasets are already beneficial, given that even limited local data can be used to estimate missing data~\citep{Milojevic-Dupont.2020b,rosser2019predicting}. 
    
    \paragraph{Fostering interoperability with other data layers}
    Many of the most impactful applications for urban sustainability come by fusing datasets on building and on other land uses, or possible satelite imagery~\citep{zhu2019understanding,silva2017urban}. When several of such datasets have been gathered by an administration, releasing them together and faciliating matching, e.g. with common `id' increases their individual relevance. A role model here is for example the French cadaster that releases openly a single structured data product, BDTOPO, which includes buildings, various infrastructures including transportation, electric grid or water pipes. When these datasets have been generated by other entities, interoperability could be mitigated by widespread implementation of best practices. 
    
    \paragraph{Leveraging the wider ecosystem}
     Non-governmental entities can have an important role in improving data quality and convergence in practices, as demonstrated in the European Union by the project INSPIRE. Initiated in 2006, this project provided technical guidelines and support to EU state members as well as a repository where user can access the data in centralised manner. Our results suggests that the EU is the region that offers the best data coverage and quality overall, although there is still substantial room for progress in all dimensions analysed. Similar initiatives at the regional or global level, for example the ISO standard 19115:2013 "Geographic Information – Metadata" may help improve the current data situation. Leveraging local partners, such as universities or other research organisations, which have demonstrated their interest and expertise~\citep{Dukai.2020}, is a promising pathway that could be followed by more governments.

\subsection{Nurturing a symbiosis with OpenStreetmap}\label{sc:symbiosis}

    \medskip
    \noindent In jurisdictions where the government maintains and releases data for public use, a question is whether the co-existence with volunteered sources such as OSM in the same geography is beneficial and under which conditions. A key open question is how open government data and OSM differ in term of user adoption~\citep{Donker.2016,Purwanto.2020}, and how potential interactions can increase adoption.

    \paragraph{Distinct advantages and drawbacks}
    
    The main advantages of OSM include a globally consistent framework with a uniform and liberal licence, a large community of contributors and increasing corporate editing~\citep{Anderson.2019}. This enables OSM to provide rapid updates (see example on Figure~\ref{fig:osm}), flexible new attributes thanks to the freeform tag-value system possibly leading to more detailed data~\citep{Knoth:2017dk}, and, a versioning approach generating historical data. OSM plays a key role in the Global South through humanitarian mapping efforts and embedding local knowledge in maps. The main drawbacks of OSM include potential quality and accuracy issues, inconsistent and scarce attributes, low budget overall for maintenance and enhancement of the data services, the fact that OSM is subject to vandalism (although there is an increasingly developing ecosystem for OSM quality control), and generally a urban/rural divide in term of completeness. However, in some cases, it has been noticed that the quality of OSM can be higher than that of government datasets~\citep{Antoniou:2015hl,Minghini.2019}.

    The main advantages of government data are their potential full completeness in term of footprints and sometimes attributes, homogeneous data collection by a single actor that can ensure quality control, and sometimes an extended set of attributes not available elsewhere thanks to the incorporation of administrative data. The main drawbacks include the potential difficulty to update and correct regularly the data, which together with the lack of historical data can hinder analyses in rapidly developing cities, the fact the data are locally limited to a jurisdiction, often available in scattered repositories with variable licenses and language barriers.

    \paragraph{Combining the best of both worlds}
    
\begin{figure}[h!]
    \centering
    \includegraphics[width=\textwidth]{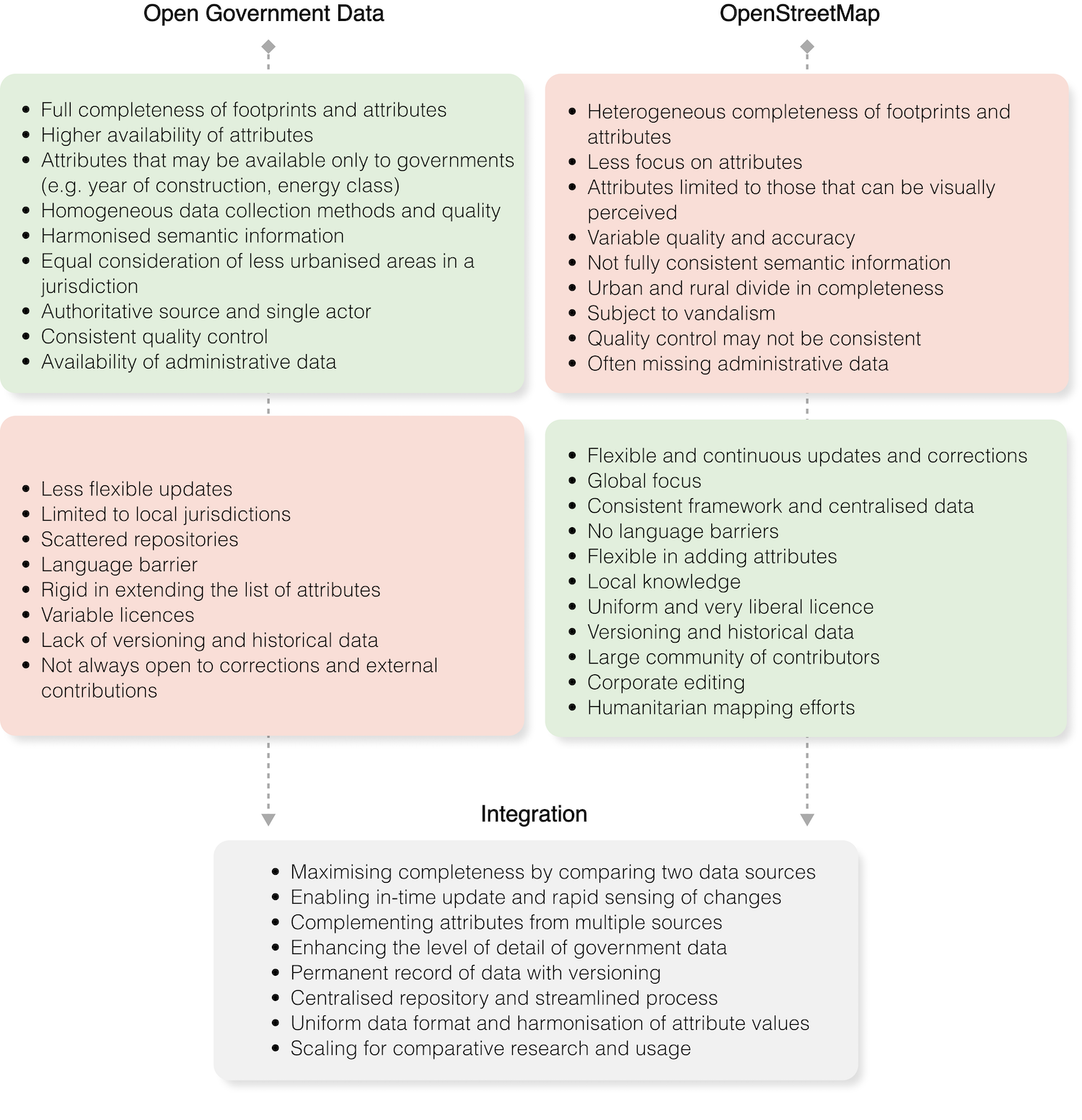}
    \caption{\textbf{Nurturing a symbiosis between open governmental data and OpenStreetMap.} Both data sources have distinct advantages (green) and drawbacks (red), and thus complement each other. There are opportunities to further integrate open government data and OSM that could result in combined datasets of higher relevance for urban sustainability and an amplification of the strengths of each data source (grey). The pros and cons are based on the majority of cases, and there may be exceptions to some of the statements in a few geographies.}
    \label{fig:ogd-osm-relation}
\end{figure}

    As both government and OSM data have distinct advantages and limitations, they do not conflict and can complement each other (Figure~\ref{fig:ogd-osm-relation}). There can be a two-way interaction between governments and VGI~\citep{Haklay.2014,Bank.2018}. 
    
    Governments can benefit from OSM and have been engaging with VGI communities. VGI may be used by governmental bodies to supplement or facilitate their operations, as VGI enables governments to capture and integrate local knowledge, which might not be considered by official acquisition processes. For example, in Indonesia, the OSM community was engaged to map buildings in areas where previously such was not available, and the resulting dataset was adopted by the government to produce scenarios for different disasters and offer a tool to develop contingency plans~\citep{Haklay.2014,Bank.2018}. It may also be useful here to differentiate across governments' level of development in term of data acquisition and maintenance. In cases where governments have highly-constrained means and that a VGI community is emerging, engaging with and investing in VGI may enable to facilitate government operations at a lower cost while generating some other benefits, e.g.\ related to education. In cases where both OSM and government have a high coverage, interactions may enable to shape how to develop future data products that of high value for the broader society.   
    
    Availability of open government data also seems beneficial to OSM. To illustrate this point, we compared the coverage of OSM and open government data in the European Union (see Supplementary Information). Countries that have highest coverage in OSM tend to have open government data available at the country scale and countries that have the lowest coverage in OSM have no open access to government data. There are rare cases of either high OSM coverage with low official data, possibly due to a particularly active community, and low OSM coverage with high authoritative data, perhaps due to government data being only newly released and not yet integrated into OSM. Another benefit is that using government data as ground truth is the most common way to carry out quality checks of OSM~\citep{Senaratne:2017bj,Yan:2020dp,Jacobs.2020}.

    Ideally, open government data and OSM could be integrated in one centralised repository, providing multiple benefits.
    First, this combination should contribute towards maximising completeness of geometries and attributes, using the higher level of detail available in case of matches. Second, it should enable rapid updates with permanent record of data accompanied by versioning. Third, there should be a streamlined process with uniform data format and harmonisation of attribute values. Such initiative could be undertaken at different scales and could result in higher data availability and quality both for local and comparative regional or global research.

In conclusion, although open government data have become increasingly available and accessible globally, high-quality data remains mostly limited to high-income countries and main urban areas. Data currently available provide often incomplete or basic snapshots of building stocks. There are large data gaps (e.g.\ no buildings at all for many areas or little information on building attributes) that may be filled by further releases, VGI, or ML-based data generation approaches, and there is a high potential of morphing into integrated data. Large disparities in practices among local and national governments worldwide in terms of several aspects such as update frequency, modelling guidelines, and metadata availability require further government efforts to provide a timely and needed empirical basis for transforming the building stock globally. While it has a limited geographic coverage in comparison to OpenStreetMap, its crowdsourced counterpart, government data usually reign supreme where they are available, often providing the level of quality that is highest among datasets from other parties; but their quality should not be taken for granted, and in some instances there may be better options available. This overview of the trends in open government building data may serve policy-makers, researchers, and practitioners to understand the current landscape of open data on buildings and plan further actions.

\section{Methods}\label{sc:methodology}

    \subsection{Definitions and inclusion criteria}
    
    \noindent In our study, we have considered datasets on buildings that satisfy three criteria.
    
    First, we follow the open data definition (\url{https://opendefinition.org/od/2.1/en/}) set by the Open Knowledge Foundation: `Open data and content can be freely used, modified, and shared by anyone for any purpose'.
    Consequently, data available to anyone but for a charge, and data available only for viewing (without the possibility of downloading it as a machine readable dataset) are instances that cannot be considered open and were thus excluded from consideration. 
    In addition, the dataset should be relatively easy to download, not requiring expert knowledge or complex workflows.
    
    Second, the dataset has to be created and released by a governmental authority, such as national mapping/cadastral agency, regional government, or city administration.
    We have not considered datasets that are not of official nature.
    For example, commercial releases and volunteered geoinformation are not in the focus of this research, though we draw parallels to them in the discussion of the findings.
    
    Third, the dataset must contain 2D spatial data on buildings (i.e.\ their footprints). For example, non-spatial datasets (e.g.\ spreadsheets or aggregated statistics) and point-based datasets (e.g.\ geocoded addresses) are not considered for this project due to their limited usefulness in geospatial workflows and urban studies.

    \subsection{Data acquisition}
    
    \noindent Given the global scope of this study, the discovery of data sources was one of the principal tasks. Data discovery was undertaken during the first half of 2021. 
    Instances of open government building data were discovered through different approaches: (i) searching specifically for building data using Google search engine, (ii) searching national open data portals (including several languages thanks to automated translation tools), (iii) searching international open data portals (e.g. INSPIRE), (iv) examining papers that use building data, personal experience and contacts, and (v) crowdsourcing through social media.
    A combination of these different approaches ensures a wide reach and gives a good level of confidence regarding the diversity of the datasets. The list of datasets included in this research however should not be regarded as exhaustive, but as indicative of the existing practices and spatial patterns in term of availability at the time of the search.
    
    \subsection{Extraction of information and data analysis}
    
    \noindent For each discovered dataset, we have noted the following information and metadata: download link, location, level of administrative coverage (e.g.\ city, region), year of creation and last update, and data format.
    In our method, we differentiate whether the dataset has been released according to a standard, since open data does not necessarily mean that the format is based on an international standard~\citep{Wilson.2020}.
    Further, we cross-checked the list of discovered datasets with the OSM website (\url{https://wiki.openstreetmap.org/wiki/Import/Catalogue}), which details the list of datasets integrated in OSM, to identify whether the dataset has been imported into OSM.
    
    Afterwards, each dataset was inspected to ensure its content, and analysed manually to extract further information for assessing the five dimensions of interest in this study. In case of larger areas (e.g.\ nation-wide datasets containing millions of buildings), we have analysed data samples instead of the entire dataset.
    
    First, the semantic content of each dataset was analysed.
    For each, we have derived a list of attributes pertaining to individual buildings that the dataset contains, such as year of construction and address of each building.
    We have not considered attributes that are not related to buildings or are not relevant for this work, e.g.\ internal identifier of a building, accuracy of the acquisition technique.
    We have also not considered attributes that are computed from the geometry of the footprint, e.g.\ area of the footprint, as these do not provide added value that is not already in the dataset.
    The decision to inspect the semantic content manually turned out to be instrumental, as we have identified datasets that nominally contain a set of attributes, but their actual content was empty in the dataset.
    Such attributes were not considered.
    Second, we have analysed the geometric validity of the data using the method developed in~\citep{prepair}, as geometrically invalid datasets have implications on the usability of data.
    
    Not stopping at the open data portal and relying on the metadata only, but rather inspecting the data in detail proved to be useful also to reveal issues and some particularities of datasets that have shaped the results and discussion of this study.
    For example, some datasets turned out not to be downloadable (e.g.\ broken links or empty files).
    If we could not verify at least the sample of a dataset, we have not considered it further and excluded it from our analysis.
    Further, we have identified several datasets that do not include all the buildings in their administrative extent.
    The types of such instances are diverse.
    For example, there are datasets that are thematic (e.g.\ they have only commercial or school buildings mapped), driven by authority and real estate (e.g.\ containing only buildings on public land), or those that have only buildings with a footprint larger than a threshold of considerable size, not being representative of buildings in the area.
    Further, some datasets have partial coverage as they have an indicative purpose, e.g.\ serving as a sample dataset rather than a complete one.
    Finally, there are jurisdictions in which data is released gradually as it is being collected as part of a large effort with a long timeline (e.g.\ nation-wide dataset with data released gradually by subdivisions as they are mapped).
    As these datasets may still be found useful for some spatial analyses, and may hint at the existence of a full dataset that is yet to be released openly, we have included them in our analysis.

    \subsection{Methodology for the assessment of the datasets}
    
    \noindent We assessed the state-of-the-art of open geospatial data on buildings across five main dimensions: accessibility, richness, data quality, harmonisation, and relationship with other actors. The five dimensions and their constitutive features were designed from a user perspective and based on domain knowledge. We aimed to cover the dimensions that are most important to enable both local and global usage of the data for urban sustainability research.  
    
    When possible, we took a quantitative approach by extracting key metrics supporting our assessment. In cases for which quantified information would be too difficult or too ambiguous to generate, you took a qualitative approach to describe as precisely as possible the patterns we observed. 
    
    All these aspects are one of the key contributions of our work, as they are embryonic -- they have not or have been scarcely discussed in international scientific literature hitherto, but they offer much potential.
        
        \subsubsection{Accessibility}
    
        \noindent In the first dimension, we scrutinise three aspects: existence, openness, and international visibility.
        First, we describe a qualitative description of the continents where data was found. Given the uncertainty of the exact number of datasets existing (more about this in the limitations), we decided to report these results qualitatively rather than quantitatively.
        Second, openness is examined qualitatively by understanding the ease of accessing the data, especially from a perspective of a user who is not a geospatial expert.
        Third, for international visibility, we regard if the portal and/or metadata are in English, the lingua franca of science and technology.

        \subsubsection{Richness}
        
        \noindent Richness is examined through three topics: spatial extent and completeness, presence of attributes, and level of detail.
        In the first, we can analyse quantitatively the level of the jurisdiction (e.g.\ city-scale). 
        By analysing the data and metadata, we also verify whether the dataset covers the entire jurisdiction.
        Second, understanding the presence of attributes is equally important as examining the geometric aspect, since they are critical in many analyses.
        For each dataset, we list the relevant attributes and examine their contents, also quantitatively.
        Last, the level of detail looks qualitatively into both the geospatial and attribute content: how detailed they are respectively in geometrical and semantic terms.
    
        \subsubsection{Quality}
        
        \noindent The three constitutive aspects of quality are all investigated quantitatively.
        First, the availability of essential metadata: for each dataset, we note whether it is unambiguous when the dataset was created and how frequently it is updated, together with the licence, a key piece of information relevant for the usability aspect.
        In the case of datasets that are updated periodically, we explore whether historical datasets are available as well.
        Second, frequency of updates is an aspect that stems from the first one, focusing on the actual date of the last update.
        Third, we inspected the validity of geometries of each dataset using the method and software of \citep{prepair}.
        
        \subsubsection{Harmonisation}
        
        \noindent In the fourth dimension --- harmonisation --- we seek to understand three aspects: standards, data formats, and definition of buildings.
        First, we examine metadata to understand whether the dataset follows an international standard for encoding building information (qualitative approach).
        Second, we list the format(s) in which data is disseminated (quantitative approach).
        Third, we seek to interpret qualitatively the definition of a building from metadata, e.g.\ whether there is a size criteria or real estate definition.
        
        \subsubsection{Relationship with other actors}

        \noindent The final dimension, fully qualitative, endeavours to understand the role official data has in the developing and increasingly saturated offering of building information.
        First, we examine the cross-pollination and relationship with OpenStreetMap, by understanding previous and current practices, and future plans.
        We also compare both datasets in several locations.
        As part of the work, we analyse a website with all the datasets that have been imported by the OSM community, together with the current progress and future plans.
        Second, we seek to examine the instances of government data curated by a private company, by understanding the lineage and possible multiplicity of repositories.
        Third, involvement of universities is an equally important aspect, which has been examined by analysing publications that use open government building data.

    \subsection{Publicly available inventory}
    
    The list of identified datasets together with their links and a map is available at the website of the Urban Analytics Lab at the National University of Singapore (https://ual.sg/project/ogbd/).
    We enable users of the index to suggest new datasets and report errors.    

    \subsection{Limitations}
    
    \noindent The list of identified datasets globally, albeit the first and largest to date to the extent of our knowledge, is nevertheless likely not a complete one, mostly due to language barriers and the imperfect reach of our methodology.
    While we have used a multi-pronged and systematic method to source datasets, and we have included all datasets that came to our attention, it is possible that some were missed.
    Thus, our analysis is based on insights derived from a subset rather than all datasets that may be available in reality.
    Further, it is possible that our exploration approach has resulted in some biases (e.g.\ towards sources in the languages spoken by the authors and their contacts) potentially skewing our analysis and discussion. While we acknowledge the possibility of such biases, we have tried our best to mitigate them by reaching out to local experts on every continents and engaging social media. Given the large extent of our index, we believe that a potential omission of some datasets does not have a substantial impact on our overall results, and that the emergence of new datasets would not substantially change the key results and discussion points.
    
    Another limitation of our work is the likelihood of potential errors in the metadata, which might be caused by mistakenly identifying an older source of data (e.g.\ previous instances of datasets may have remained online and on different websites, despite a newer version being available), or misreading the documentation, especially if automated translation tools have been used.
    However, as with the previous point we expose, it is unlikely that the key results of our study have been affected significantly by such inconsistencies.

\section*{Data Availability}
\noindent The data generated and analysed during this study is available at \url{https://doi.org/10.25540/YC94-A3X2}.

\begin{table}
\caption{Summary of the analysis of open government geospatial datasets on buildings.}
\label{tab:summary}
\scriptsize
\begin{adjustwidth}{-1.2in}{+0in} 
\renewcommand{\arraystretch}{1.3}
\begin{tabular}{|p{2cm}p{2.8cm}p{4.2cm}p{4cm}p{4cm}|}

\toprule
\textbf{Dimension}                       & \textbf{Feature}                                    & \textbf{Current status}                                                                                                                  & \textbf{Implications on usage for urban sustainability research}                                                                                    & \textbf{Recommendations on best practices}                                                                                                  \\\midrule
\textbf{Accessibility}                   & Existence                                  & Availability is mostly restricted to  high-income countries and main urban areas                                                   & The Global South, fast urbanizing regions are almost not represented                                                                       & Increasing coverage, providing at least footprints                                                                      \\
                                & Openness                                   & A majority of datasets are easily accessible without cost or registration, but sometimes APIs are complex           & Many datasets can be used for open science research; possibility to enrich and republish the data                  & Following the open data approach from Open Knowledge Foundation                                                                                                        \\
                                & International visibility                   & Datasets may be difficult to locate; e.g.\ for some, the title does not even contain the word `building' & Many datasets remain difficult to access for an international audience                                              & Listing the dataset on national and international open data platforms; using keywords in several languages \\
\textbf{Richness}                        & Spatial scale                              & Mostly city-level, some region and country-level datasets; several datasets are just samples of an area             & Usage for large-scale studies is limited as there are large data gaps, even for best-covered continents            & When a national cadastral service exists, publishing country-level data e.g.\ Spain or France               \\
                                & Presence of attributes                     & Attributes are largely unavailable, e.g.\  type 34\%, height 21\%, age 16\%                       & Limited information available about the properties of the building stock                                      & Focusing on collecting most important attributes as minimum; linking with other data, e.g.\ land registry   \\
                                & Level of detail                            & Most datasets provide polygons of the outline of the building, few provide just the point coordinates      & Most datasets are usable for urban morphology studies                                                              & Providing explicit building geometries in 2D or 3D                                                                    \\

\textbf{Quality}                         & \raggedright Availability of essential metadata         & Year of update and acquisition method often present, but licenses are often difficult to access              & Some basic information to assess the conditions of usage may be missing                                            & Having at other basic metadata, such as clear licence, quality, and standard                                                                                    \\
                                & Frequency of updates                       & Large fraction of datasets are up-to-date (48\% are no more than a year old) but temporal archives are rare     & Datasets are suitable to study the current state of cities but not the evolution of urbanisation processes         & Archiving previous versions instead of overwriting them; enabling versioning                                                     \\
                                                                & Validity of geometries                     & Negligible share of invalid geometries, but this does not guarantee positional accuracy        & 2D datasets are virtually free of errors; in contrast, 3D datasets often have \textgreater{}50\% invalid geometries &                                                                                \\

\textbf{Harmonization}                   & Standard identifiable                      & International standards are not a common practice; in the EU there are progresses with INSPIRE          & For comparative studies, users need to verify each dataset with respect to modelling guidelines, licenses...                                                                                                                & Partnerships with non-governmental organisations to establish new standards                                                                                                                                                           \\
                                & Format                       & Proprietary Shapefile is the most common format (86\%); various others are also used, incl.\ custom formats                                                                         & Potential challenges for non-expert users                                                                & Providing entirely open formats, e.g.\ GeoJSON     
                                                       \\
                                & Definition of a building                   & The definition of a building is not standardised around the world                                       & Potential impact on large-scale models (noise, etc.) and (comparative) analyses                                                  & International definition of a building and standardisation \\
\raggedright \textbf{Relationships with other actors} & \raggedright Integration in OpenStreetMap               & Only 25\% of the datasets have been integrated in OpenStreetMap                                          & Limited share of government data to be found on OSM, need to gather data from individual websites                  & Clearer licences, rapid integration, and transportation of semantic information                                                         \\
                                & Curation or distribution by private companies & Relatively marginal practice                                                                       & Vendor lock-in                                                                                                                & Archiving data by non-profit stakeholders and immediate integration in OSM                                                                                                        \\
                                & \raggedright Involvement of universities                & Examples of quality control, enrichment, and linkage with other datasets exist         & Further guarantees for scientific usage of the data                                                                & Building relationships with local universities, including funding                                          \\ \bottomrule
\end{tabular}
\end{adjustwidth}
\end{table}

\section*{Acknowledgements}

\noindent We are grateful to all contributors who have pointed out locations with authoritative building data.
We thank the members of the NUS Urban Analytics Lab for the feedback, April Zhu for the design contribution, and Marco Minghini and Giacomo Martirano from the European Union's Joint Research Centre for the insightful discussions about open government data in the EU.
This research is part of the project Large-scale 3D Geospatial Data for Urban Analytics, which is supported by the National University of Singapore under the Start Up Grant R-295-000-171-133.

\section*{Author Contributions}
\noindent Study conception: FB; Study design: FB and NMD; Data collection: LZXC; Analysis and interpretation of results: FB, NMD, LZXC; Paper preparation: FB, NMD; FC revised and provided critically important content; All authors contributed to the manuscript and approved the final version for submission.

\section*{Competing Interests}

\noindent The authors declare no competing interests.

\section{Supplementary information}

\begin{table}
\begin{tabular}{ | l | c | l | }
\hline
	\textbf{Country} & \textbf{Coverage OSM} & \textbf{Availability open gov.\ data }  \\ \hline
	Austria & \textcolor{green}{0.92} & \textcolor{orange}{partial, no restrictions}  \\ 
	Belgium & \textcolor{green}{0.69} & \textcolor{green}{complete, no restrictions} \\ 
	Bulgaria & \textcolor{red}{0.10} & \textcolor{red}{not available}  \\ 
	Croatia & \textcolor{orange}{0.37} & \textcolor{red}{restricted access}  \\ 
	Cyprus & \textcolor{red}{0.11} & \textcolor{red}{not available}  \\ 
	Czech republic & \textcolor{green}{0.95} & \textcolor{green}{complete, no restrictions} \\ 
	Denmark & \textcolor{green}{0.92} & \textcolor{green}{complete, no restrictions} \\
	Estonia & \textcolor{orange}{0.58} & \textcolor{green}{complete, no restrictions}  \\ 
	Finland & \textcolor{orange}{0.56} & \textcolor{green}{complete, no restrictions}  \\ 
	France & \textcolor{green}{1.07} & \textcolor{green}{complete, no restrictions}  \\ 
	Germany & \textcolor{green}{0.78} & \textcolor{orange}{partial, no restrictions}  \\ 
	Greece & \textcolor{red}{0.09} & \textcolor{red}{not available}   \\ 
	Hungary & \textcolor{orange}{0.30} & unsure   \\ 
	Ireland & \textcolor{green}{0.70} & \textcolor{red}{restricted access}   \\ 
	Italy & \textcolor{orange}{0.38} & \textcolor{orange}{partial, no restrictions}   \\ 
	Latvia & \textcolor{orange}{0.40} & \textcolor{orange}{on request}  \\ 
	Lithuania & \textcolor{green}{0.81} & \textcolor{green}{complete, no restrictions}   \\ 
	Luxembourg & \textcolor{orange}{0.63} & \textcolor{green}{complete, no restrictions}   \\ 
	Malta & \textcolor{red}{0.07} & \textcolor{green}{complete, no restrictions}   \\
	Netherlands & \textcolor{green}{1.11} & \textcolor{green}{complete, no restrictions}  \\ 
	Poland & \textcolor{green}{0.87} & \textcolor{green}{complete, no restrictions}  \\ 
	Portugal & \textcolor{red}{0.16} & \textcolor{red}{restricted access}   \\ 
	Romania & \textcolor{red}{0.13} & \textcolor{red}{restricted access}   \\ 
	Slovakia & \textcolor{green}{1.12} & \textcolor{green}{complete, no restrictions}   \\ 
	Slovenia & \textcolor{orange}{0.51} & \textcolor{green}{complete, no restrictions}   \\ 
	Spain & \textcolor{red}{0.13} & \textcolor{green}{complete, no restrictions}  \\ 
	Sweden & \textcolor{orange}{0.44} & \textcolor{red}{restricted access}  \\ \hline
\end{tabular}
\caption{\textbf{Comparison of building data available in OpenStreetMap (OSM) and open government data in the European Union.} The coverage ratio for OpenStreetMap is computed by dividing the number of buildings downloaded on in February 2021 by the number of buildings reported in EUROSTAT. The availability of open government data is based on self-reported information by governments in the INSPIRE geoportal. Note that for some datasets indicated here as complete, we were not able to download the data e.g. due to technical problems with the download servers, and thus did not include them in our main analysis. It is however assumed that users with knowledge of the local context and technical background, such as OpenStreetMap contributors, may be able to access these data and make use of it.}
\label{osm-gov-table}
\end{table}

\end{document}